\documentclass[12pt,english]{article}
\usepackage[T1]{fontenc}
\usepackage[latin1]{inputenc}
\usepackage{geometry}
\usepackage{longtable}
\usepackage{graphicx}
\usepackage{setspace}
\usepackage{amssymb}

\makeatletter

\providecommand{\tabularnewline}{\\}

\date{}
\usepackage{graphicx}
\def\be{\begin{equation}}
\def\ee{\end{equation}}
\def\bea{\begin{eqnarray}}
\def\eea{\end{eqnarray}}

\usepackage{babel}
\makeatother
\begin{document}

\title{{\Large Spin Two Glueball Mass and Glueball Regge Trajectory from
Supergravity }}

\author{\textsf{\textsc{\vspace{0.5in}}}}

\author{\textsf{\textsc{\normalsize Xavier Amador}}\textsf{\textsc{}}%
\footnote{\texttt{xamador@fis.cinvestav.mx}%
} \textsf{\textsc{~}}\textsf{\textsc{\normalsize and ~Elena Caceres}}\textsf{\textsc{}}%
\footnote{\texttt{caceres@fis.cinvestav.mx}%
}}

\maketitle
\begin{center}{\center{\it Departamento de F\'\i sica, Centro de Investigaci\'on y de Estudios Avanzados del IPN, CINVESTAV,

Apdo. Postal 14-740, 07000 M\'exico, D.F, M\'exico.}

\vspace{1.5in}}

\textsl{Dedicated to the memory of Iciar Isusi}.\end{center}

\vspace{0.5in}

\begin{center}ABSTRACT\end{center}

We calculate the mass of the lowest lying spin two glueball in $\mathcal{N}=1$
super Yang-Mills from the dual Klebanov-Strassler background. We show
that the glueball Regge trajectory obtained is linear; the $2^{++}$
, $1^{--}$ and $0^{++}$ states lie on a line of slope $0.23 \ (\frac{{g_s}^2 M}{\epsilon^{4/3}})$. We also compare mass ratios with lattice data and find agreement
within one standard deviation.

\vfill\eject

\begin{onehalfspace}

\section{\textsc{\large Introduction and summary} {\large \label{sec:Introduction}}}
\end{onehalfspace}

The formulation of a gauge/string duality \cite{Maldacena:1998re,Gubser:1998bc}
provides a new framework to study confining phenomena. Several supergravity
duals to confining gauge theories are known \cite{Polchinski:2000uf,Klebanov:2000hb,Maldacena:2000yy}.
We will focus on the Klebanov-Strassler (KS) IIB supergravity solution
\cite{Klebanov:2000hb} which describes N regular and M fractional
D3 branes on a deformed conifold space . This background is dual to
a cascading gauge theory with $SU(M+N)\times SU(N)$ gauge group in
the ultraviolet and flows, through a series of Seiberg-Witten dualities,
to $SU(M)$ $\mathcal{N=}1$ super Yang-Mills in the infrared. The
glueball spectrum for $0^{++}$ and $1^{--}$ in the KS background
was obtained in \cite{Caceres:2000qe,Krasnitz:2000ir}. Recently the
meson spectrum was investigated in \cite{Sakai:2003wu}. In this paper
we calculate the spin 2 glueball mass. The interest is two-fold; it
provides one more point in the Regge trajectory and allows us to compare
mass ratios with lattice results.

Regge theory successfully describes a large quantity of experimental
data\cite{Collins:1977co,Donnachie:2002bln}. It predicts that composite
particles of a given set of quantum numbers, different only in their
spin, will lie on a linear trajectory \[
J=\alpha_{0}+\alpha^{'}t\]
where $J$ is the spin and $t$ is the mass squared. Regge theory
treats the strong interaction as the exchange of a complete trajectory
of particles. With the inclusion of a soft pomeron, this approach
successfully describes the high energy scattering of hadrons. Understanding
Regge theory from first principles is undoubtedly a remarkable challenge.
In \cite{Pando-Zayaz:2003} it was noted that the glueball masses
found in \cite{Caceres:2000qe} provide an impressive numerical match
for the slope of the soft pomeron trajectory. The value obtained for
the Regge slope was calculated with two points ; $0^{++}$and $1^{--}$.
In  this note we obtain the mass of the $2^{++}$ glueball which provides
the next point on the Regge trajectory. The eigenvalue found for the lowest lying spin two glueball is $m^{2}(2^{++})=18.33 $ measured in units of the conifold deformation, $\frac{\epsilon^{4/3}}{g_{s}^{2}M^{2}\alpha^{'^{2}}}$.
We find that on a Chew-Fraustchi\cite{Chew:1961ev}
plot (J vs. $m^{2}$) this value of the $2^{++}$ mass lies in the same line as the previously
found $0^{++}$and $1^{--}$. Thus, we show that the KS background
predicts a \textit{linear} glueball trajectory, $J=-2.29+0.23\,\, t$.
Unlike the scenario where glueball masses are identified with classical
solutions of folded strings, here there is no \textit{a priori} reason
for the Regge trajectory to be linear. That it turns out to be so
is remarkable.

Finding the mass of the spin two glueball also completes the spectrum
obtained by C\'aceres and Hern\'andez \cite{Caceres:2000qe} and
allows us to compare glueball mass ratios with lattice results. Lattice
data for $D=4$ is not as abundant as for $D=3$. Lucini and Teper
explored the $N\rightarrow\infty$ limit of $SU(N)$ Yang-Mills theory
in four dimensions \cite{Lucini:2001ej}. We present their results
in Table 1 and show that the agreement with supergravity results obtained
from the KS background is within one standard deviation. It is interesting
to include in the comparison lattice results for QCD in $D=4.$ Recently
Morningstar and Peardon\cite{Morningstar:2003ew} improved on their
previous results \cite{Morningstar:1999rf} for the glueball spectrum
of QCD. Using state-of-the-art techniques they determined the $0^{++}$and
$0^{++*}$masses in a more reliable way; their results are also shown
in Table 1.

It is worth noting that the supergravity theory we are considering
is dual to an embedding of $\mathcal{N=}1$ SYM into IIB string theory,
the agreement with non-supersymmetric $SU(\infty)$ lattice results
seems to indicate that, at least as glueball mass ratios are concerned,
they are in the same universality class.

\begin{table}
\begin{center}\begin{tabular}{|c||c|c||c|}
\hline 
&
KS model&
Lattice $SU(\infty)$&
Lattice $QCD_{4}$\tabularnewline
\hline
\hline 
$m(0^{++*})/m(0^{++})$&
$1.84$&
$1.91(17)$&
$1.79(6)$\footnotemark\tabularnewline
\hline 
$m(2^{++})/m(0^{++})$&
$1.37$&
$1.46(11)$&
$1.39(4)$\footnotemark\tabularnewline
\hline
\end{tabular}\end{center}

\caption{Glueball mass ratios calculated in the KS model and in two lattice
simulations; $SU(\infty)$\cite{Lucini:2001ej} and $SU(3)$ \cite{Morningstar:2003ew,Morningstar:1999rf}.}
\end{table}

\addtocounter{footnote}{-1}

\footnotetext{Preliminary result \ \cite{Morningstar:2003ew}}

\stepcounter{footnote}\footnotetext{\cite{Morningstar:1999rf}}To calculate the $2^{++}$ground state we solve the relevant eigenvalue
problem numerically using a {}``shooting'' algorithm. This method
is more accurate than W.K.B approximation for low-lying states but
is sensitive to the choice of initial guess. Since the equation to
solve is numerically delicate we first solve the problem using W.K.B.
We then use the eigenvalue found with W.K.B as initial guess in the
shooting method. 

\section{\textsc{\large The Klebanov-Strassler supergravity dual of $\mathcal{N}$=1
Yang-Mills}{\large \label{sec:KSsugra}}}

The Klebanov-Strassler IIB supergravity solution \cite{Klebanov:2000hb}
describes N regular and M fractional D3 branes on a deformed conifold.
In the deformed conifold the conical singularity is removed through
a blow-up of the $S^{3}$ of the base. The KS solution is rich in
interesting physical phenomena; exhibits confinement, chiral symmetry
breaking, dimensional transmutation, domain walls etc. Here we will
review some aspects of the KS background necessary for the next sections
(for a more detailed account of the KS background see \cite{Herzog:2002}
). 

The KS solution consists of a warped deformed conifold transverse
space and non-vanishing 3-form and 5-form fluxes. The ten-dimensional
metric takes the form,

\begin{equation}
ds_{10}^{2}=h^{-1/2}(\tau)dx_{n}dx_{n}+h^{1/2}(\tau)ds_{6}^{2}\label{eq:tendimspace}\end{equation}
 where \[
h(\tau)=(g_{s}M\alpha')^{2}2^{2/3}\epsilon^{-8/3}I(\tau),\]
and\[
I(\tau)\equiv\int_{\tau}^{\infty}dx\ \frac{(x\coth x-1)(\sinh(2x)-2x)^{1/3}}{\sinh^{2}x}\]
is a dimensionless function. The metric of the deformed conifold in
its diagonal basis is,\bea {ds_6}^2=\frac{1}{2}\epsilon^{4/3}K(\tau)\left[\frac{1}{3K^3(\tau)}(d^2\tau+(g^5)^2)\right. &+&\cosh^2(\frac{\tau}{2})[(g^3)^2 +(g^4)^2]\nonumber \\&+& \left.\sinh^2(\frac{\tau}{2})[(g^1)^2 +(g^2)^2]\rule{0cm}{0.6cm}\right]\label{defconifoldmetric}\eea
where\[
K(\tau)=\frac{(\sinh(2\tau)-2\tau)^{(1/3)}}{2^{1/3}\sinh(\tau)}.\]

The self-dual 5-form field strength may be decomposed as  $\tilde F_5 = {\cal F}_5 + \star {\cal F}_5$
where

\be \star{\cal F}_5 = 4 g_s M^2 (\alpha')^2 \varepsilon^{-8/3} dx^0\wedge dx^1\wedge dx^2\wedge dx^3 \wedge d\tau {\ell(\tau)\over K(\tau)^2 h(\tau)^2 \sinh^2 (\tau)}\ , \ee 
and 

\be {\cal F}_5 = B_2\wedge F_3 = {g_s M^2 (\alpha')^2\over 4} \ell(\tau) g^1\wedge g^2\wedge g^3\wedge g^4\wedge g^5\ .\ee 

The complex three form is $G_3 = F_3 + \frac{i}{g_s} H_3$ with

\be F_3 ={M\alpha'\over 2} \left \{g^5\wedge g^3\wedge g^4 + d [ F(\tau) (g^1\wedge g^3 + g^2\wedge g^4)]\right \}  \ee

 \begin{eqnarray} H_3 = dB_2 &=& {g_s M \alpha'\over 2} \bigg[ d\tau\wedge (f' g^1\wedge g^2 +  k' g^3\wedge g^4) \nonumber \\ && \left. + {1\over 2} (k-f) g^5\wedge (g^1\wedge g^3 + g^2\wedge g^4) \right]\ . \end{eqnarray} 

The functions defining the three and five form are \bea F(\tau) &=& {\sinh \tau -\tau\over 2\sinh\tau}\ , \nonumber \\ f(\tau) &=& {\tau\coth\tau - 1\over 2\sinh\tau}(\cosh\tau-1) \ , \nonumber \\ k(\tau) &=& {\tau\coth\tau - 1\over 2\sinh\tau}(\cosh\tau+1) \, \nonumber \\l(\tau) &=& \frac{\tau\coth\tau -1}{4\sinh^2\tau}(\sinh 2\tau -2\tau).\eea We
are interested in the infrared physics described by the KS background.
Note that $I(\tau)$ is non singular at the tip of the conifold, $I(\tau\rightarrow0)=a_{0}+O(\tau^{2})$
with $a_{0}\approx0.71$. Thus, near $\tau=0$ the ten dimensional
geometry is $\mathbb{R}^{3,1}$times the deformed conifold,

\begin{eqnarray} ds_{10}^2  &\rightarrow&  { \varepsilon^{4/3}\over 2^{1/3} a_0^{1/2} g_s M\alpha'} dx_n dx_n  + a_0^{1/2} 6^{-1/3} (g_s M\alpha') \bigg \{ {1\over 2} d\tau^2  + {1\over 2} (g^5)^2 \nonumber \\ && + (g^3)^2 + (g^4)^2      + {1\over 4}\tau^2 [(g^1)^2 + (g^2)^2] \bigg \} \ . \label{apex} \end{eqnarray} 

As $\tau\rightarrow0$ the $S^{3}$at the base remains finite while
the $S^{2}$ shrinks to zero. The parameter $\epsilon^{2/3}$ has
dimensions of length and measures the deformation of the conifold.
From the IR metric (\ref{apex}) it is clear that $\epsilon^{2/3}$also
sets the dynamically generated 4-d mass scale 

\[
\frac{\epsilon^{2/3}}{\alpha'\sqrt{g_{s}M}}.\]
The glueball masses scale as $m_{glueball}\sim\frac{{\epsilon^{2/3}}}{\alpha^{'}g_{s}M}$.

\section{\textsc{\large Spin Two Glueball}{\large \label{sec:Sugraequations}}}

Consider the metric, 

\[
g_{MN}=g_{MN}^{KS}+h_{MN}\]
 where $g_{MN}^{KS}$ is the Klebanov-Strassler background metric
(eq. \ref{eq:tendimspace}) and $h_{MN}$ denotes fluctuations around
this background. In order to simplify the equations of motion it is
standard procedure to make an expansion in harmonics on the angular
part of the transverse space. In the present case we are interested
in infrared phenomena \textit{i.e.} in the $\tau\rightarrow0$ region.
In this region the angular part of the deformed conifold behaves as
an $S^{3}$ that remains finite -with radius of order $g_{s}M$ at
$\tau=0-$ and an $S^{2}$ which shrinks like $\tau^{2}$ . Thus,
for small $\tau$ it is appropriate to expand in spherical harmonics
on the $S^{3}$. Note that as in \cite{Caceres:2000qe}, it is only
for the expansion in harmonics that we consider the base to be $S^{3}$;
to solve the equations of motion we will certainly consider the full
deformed conifold background given by (\ref{eq:tendimspace}). 

After introducing the expansion in harmonics in the IIB supergravity
equations and keeping in mind that we are interested in fluctuations
on the four dimensional space transverse to the deformed conifold
we find that the linearized equation for the fluctuations is\begin{eqnarray}
-\frac{1}{2}\nabla^{\lambda}\nabla_{\lambda}h_{ij}(\tau,\bar{x})\,+\frac{1}{2}\nabla^{l}\nabla_{i}h_{lj}(\tau,\bar{x})\,+\frac{1}{2}\nabla^{l}\nabla_{j}h_{li}(\tau,\bar{x})=\,\,\,\,\,\,\,\,\,\,\,\,\,\,\,\,\,\,\,\,\,\,\,\,\,\,\,\,\,\,\,\nonumber \\
\left(\,\frac{g_{s}^{2}}{96}\,\left(\frac{1}{5}\star\mathcal{F}_{5}^{KS}\cdot\star\mathcal{F}_{5}^{KS}\right)\,-\left.\frac{g_{s}^{2}}{48}H_{3}^{KS}\cdot H_{3}^{KS}-\frac{1}{48}F_{3}^{KS}\cdot F_{3}^{KS}\right)h_{ij}(\tau,\bar{x})\right.\label{eq:einsteinlinear}\\
\nonumber \end{eqnarray}
 where the covariant derivative is with respect to the full KS background
(\ref{eq:tendimspace}). Expanding in plane waves, \[
g_{ij}(\tau,\bar{x})=h_{ij}(\tau)e^{ikx},\]
a mode of momentum $k$ has a mass $k^{2}=-m^{2}.$ The spin 2 representation
of the $SO(3)$ symmetry in $x_{2},x_{3},x_{4}$ is a symmetric traceless
tensor. Choosing a gauge $g_{1i}(\tau,\bar{x})=0$, the fluctuation
$h_{ij}(\tau)$ ( $i,j=1,2,3,4$) has five independent components
corresponding to the five polarizations of the $2^{++}.$ As expected,
all five satisfy the same equation of motion and are thus degenerate.
Denoting $g(\tau)\equiv h_{22}(\tau)=h_{33}(\tau)=h_{23}(\tau)=h_{24}(\tau)=h_{34}(\tau)$
we obtain (see Appendix for details) from (\ref{eq:einsteinlinear}),

\begin{equation}
\frac{d^{2}}{d\tau^{2}}g\left(\tau\right)+A\left(\tau\right)\frac{d}{d\tau}g\left(\tau\right)+\left(B\left(\tau\right)-\frac{g_{s}^{2}\alpha^{2}M^{2}}{2^{1/3}\epsilon^{4/3}}I(\tau)G_{55}(\tau)k^{2}\right)g\left(\tau\right)=0\label{eq:sugraeq}\end{equation}

where, \[
A(\tau)=\frac{d\ln(G_{99}(\tau))}{d\tau}+\frac{d\ln(G_{77}(\tau))}{d\tau}+\frac{d\ln I(\tau)}{d\tau}\]
 and

\begin{eqnarray*}
B(\tau) & = & \frac{-2^{(1/3)}}{8I(\tau)G_{77}(\tau)^{2}}(1-F(\tau))^{2}-(\frac{d}{d\tau}k(\tau))^{2}-\frac{2^{1/3}}{16G_{77}(\tau)G_{99}(\tau)}(k(\tau)-f(\tau))^{2}+\\
 &  & 4(\frac{d}{d\tau}F(\tau))^{2}-\frac{2^{1/3}}{8I(\tau)G_{99}(\tau)^{2}}((\frac{d}{d\tau}f(\tau))^{2}+F(\tau)^{2})\\
 &  & +\frac{1}{4I(\tau)^{2}}\left(\frac{dI(\tau)}{d\tau}\right)^{2}-\frac{2^{5/3}l(\tau)^{2}}{I(\tau)^{2}K(\tau)^{4}(\cosh^{2}\tau-1)^{2}}\\
\end{eqnarray*}

The functions $F(\tau),\,\, k(\tau),\,\, f(\tau),\,\, K(\tau),\,\, l(\tau)$
are defined in section (\ref{sec:KSsugra}). $G_{77}(\tau)$and $G_{99}(\tau)$
are redefinitions of the background metric $g_{MN}^{KS}(\tau)$ such
that $g_{ii}^{KS}(\tau)=h(\tau)^{-1/2}G_{ii}(\tau)$ for $i=1...4$
and $g_{\mu\mu}^{KS}(\tau)=h(\tau)^{1/2}\epsilon^{4/3}G_{\mu\mu}(\tau)$
for $\mu=5...10$ , they do not contain dimensionfull quantities.

\subsection{\textsc{W.K.B approximation}}

Making a field redefinition, \[
g(\tau)=\psi(\tau)e^{-\frac{1}{2}\int A(\tau)d\tau}\]
we can write equation (\ref{eq:sugraeq}) in Schr\'odinger form,

\[
\frac{d^{2}\psi(\tau)}{d\tau}-V(\tau)\psi(\tau)=0\]
 \begin{equation}
V(\tau)=\frac{1}{4}A(\tau)^{2}+\frac{1}{2}\frac{dA(\tau)}{d\tau}-(-\frac{g_{s}^{2}\alpha^{2}M^{2}}{2^{1/3}\epsilon^{4/3}}I(\tau)G_{55}(\tau)k^{2}+B\left(\tau\right))\label{eq:schpot}\end{equation}
 Near $\tau=0$ the potential (\ref{eq:schpot}) behaves as\[
V(\tau\rightarrow0)\sim2.699-(0.71)(0.218)m^{2}\]

\begin{figure}
\noindent \begin{center}\includegraphics[%
  bb=0bp 100bp 800bp 550bp,
  scale=0.4]{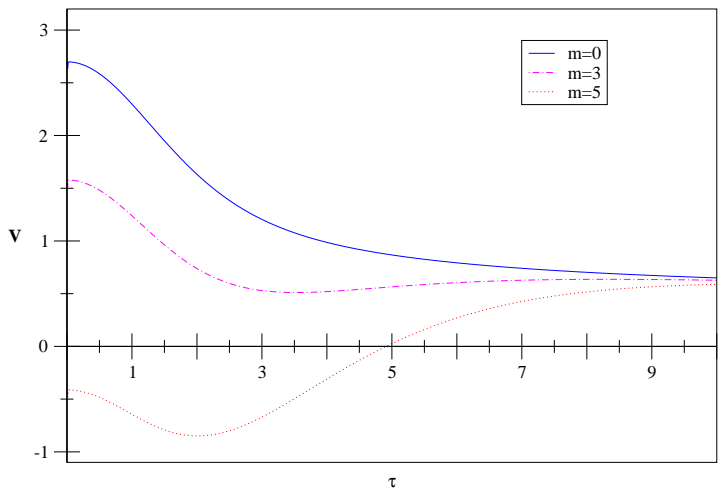}\end{center}

\caption{Potential V($\tau)$ for different values of $-k^{2}=m^{2}$}
\end{figure}

Thus, for sufficiently large $m^{2}$ the potential is negative and
there will be normalizable modes, Figure 1. The WKB approximation
consists in demanding that at the turning point $\tau_{o}$ , the
exponential solution on one side matches the oscillatory solution
on the other. The renormalizable modes are determined by the transcendental
equation\[
\int_{0}^{\tau_{0}(m^{2})}{\sqrt{|V(x)|}}dx=\frac{3\pi}{4}+(n-1)\pi\]
 where $n=1,2,3...$and $\tau_{0}$ is a root of $V(\tau_{0})=0.$ Solving
this equation numerically we find the first eigenvalue \[
m^2_{0}=21.90\]
 where the mass is measured in units of the conifold deformation $\epsilon^{4/3}.$

\subsection{\textsc{Exact numerical solution of supergravity equations}}

Equation (\ref{eq:sugraeq}) is an eigenvalue problem that can be
solved exactly by a variety of numerical methods. The boundary condition
at infinity is found by demanding normalizability of the states. Thus,
we require that $\int|g(\tau)|^{2}dx^{10}$ converges and investigating
the asymptotic behavior of the equation we find that at infinity $g(\tau)\sim\exp(-\frac{4}{3}\tau)$.
Close to the origin we will demand the function to be smooth \textit{i.e}
$\left.\frac{dg(\tau)}{d\tau}\right|_{\tau=0}=0.$ Therefore, we want
to find the eigenvalues $k^{2}$ for which there exits a solution
of satisfying the boundary conditions discussed above. We choose to
solve this problem using a shooting technique. This method is very
accurate for low lying states but requires an initial guess for the
eigenvalue. We used as initial guess the eigenvalue found in the previous
section using a WKB approximation. The coefficients entering the equation,
$A(\tau)$and $B(\tau),$ involve combinations of hyperbolic functions
that make them extremely sensitive to accumulation of numerical error.
In order to overcome this difficulty 
we calculate  $A(\tau)$and $B(\tau)$ with 20 digits of precision .
With this technique we  find a very stable eigenvalue,\[m^{2}(2^{++})=18.33 .\]

We did not find any excited $2^{++*}$ state, it is possible  that
the excited states are too heavy for us to see them since our code was design to determine accurately the lowest mode. 

\section{\textsc{\large Spectrum and Regge slope}{\large \label{sec:SpectrumRegge}}}

The low-lying glueball masses for the KS model found by C\'aceres
and Hern\'andez \cite{Caceres:2000qe} and the one obtained in this
work are shown in the following table,

\begin{center}%
\begin{table}[h]
\begin{center}\begin{longtable}{|c||c|}
\hline
\hline 
State&
\multicolumn{1}{c|}{$Mass^{2}/\epsilon^{4/3}$}\tabularnewline
\hline
\hline
\endhead
\hline 
$0^{++}$&
9.78\tabularnewline
\hline 
$1^{--}$&
14.05\tabularnewline
\hline 
$2^{++}$&
18.33\tabularnewline
\hline
\end{longtable}\end{center}

\caption{Glueball masses for the KS model.}
\end{table}
\end{center}

The Chew-Frautschi plot for the glueball trajectory obtained with
these values is shown in Figure 2. It is remarkable that the three
states lie on a straight line. For large quantum numbers it is known
that glueballs can be identified with spinning folded closed strings.
In that approach a linear Regge trajectory is no surprise since it
is built in the formalism. But in the present framework, where we
identify masses with eigenvalues of equations of motion, there is
no \textit{a priori} reason for the eigenvalues to lie on a straight
line. The fact that it is so is remarkable. The glueball Regge trajectory
obtained from the KS model is

\[
J=-2.2+0.23\, t\]

Lattice data for the intercept of the glueball trajectory  is inconclusive. While Morningstar
and Peardon find a negative intercept \cite{Morningstar:1999rf}, Teper
finds a positive one \cite{Teper:1998kw}. If the glueball trajectory
is identified with the soft Pomeron then experimental data suggests
the intercept is positive. But we do not pretend to compare the full
Regge trajectory obtained here with experimental parameters. We do
not know how to fix the scale $\epsilon^{2/3}$ and thus, any direct
comparison is far fetched. Furthermore, as in \cite{Pando-Zayaz:2003},
quantum corrections might turn out to be crucial to modify the intercept. 

\begin{center}%
\begin{figure}[h]
\begin{center}\includegraphics[%
  clip,
  width=4in,
  keepaspectratio,
  angle=-90]{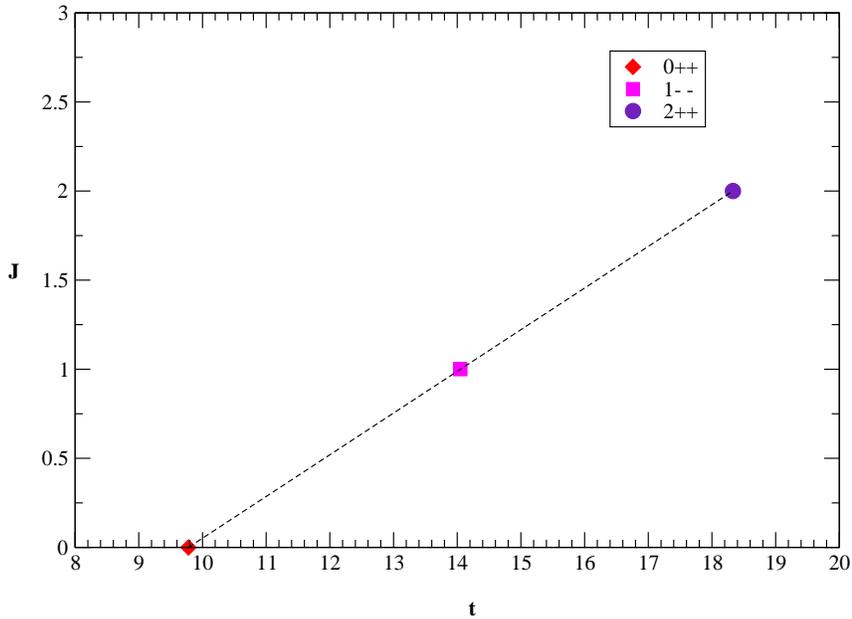}\end{center}

\caption{Glueball trajectory. The mass squared, $t$, is measured in units
of $\frac{\epsilon^{4/3}}{g_{s}^{2}\alpha'^{2}M^{2}}$.}
\end{figure}
\end{center}

\section{\textsc{\large Conclusions}{\large \label{sec:Conclusions}}}

We have calculated the mass of the spin 2 glueball in the Klebanov-Strassler
background. We find $m^{2}(2^{++})=18.33\frac{\epsilon^{4/3}}{g_{s}^{2}\alpha'^{2}M^{2}}$.
We showed that the Regge trajectory obtained is linear with a slope
of $0.23\,(\frac{g_{s}^{2}\alpha'^{2}M^{2}}{\epsilon^{4/3}})$. Comparison
of mass ratios with lattice results is in agreement within one sigma
error. The intercept of the Regge trajectory found is negative. We
expect that quantum corrections will modify the value of the intercept
and will make the trajectory non-linear. It would be interesting to
study this issue in detail.

\section*{\textsc{\large Acknowledgments}}

We are indebted to Rafael Hern\'andez for discussions and for comments
on the manuscript. We are also grateful to Leopoldo Pando-Zayas for
useful discussions and to Colin Morningstar for correspondence. The
work of E.C. is supported by Mexico's National Council of Science
and Technology (CONACyT) and that of X.A. by a graduate student fellowship
from Mexico's Office of Public Education (SEP).

\renewcommand{\theequation}{A-\arabic{equation}}

\setcounter{equation}{0}

\section*{}

\section*{Appendix A\label{sec:Appendix-A}}

In this appendix we will show some details of the derivation of (\ref{eq:sugraeq}).
After expanding in spherical harmonics the equation to be solved is,

\begin{eqnarray}
-\frac{1}{2}\nabla^{\lambda}\nabla_{\lambda}h_{ij}(\tau,\bar{x})\,+\frac{1}{2}\nabla^{l}\nabla_{i}h_{lj}(\tau,\bar{x})\,+\frac{1}{2}\nabla^{l}\nabla_{j}h_{li}(\tau,\bar{x})=\,\,\,\,\,\,\,\,\,\,\,\,\,\,\,\,\,\,\,\,\,\,\,\,\,\,\,\,\,\,\,\nonumber \\
\nonumber \\\left(\,\frac{g_{s}^{2}}{96}\,\left(\frac{1}{5}\mathcal{\star F}_{5}^{KS}\cdot\mathcal{\star F}_{5}^{KS}\right)\,-\left.\frac{g_{s}^{2}}{48}H_{3}^{KS}\cdot H_{3}^{KS}-\frac{1}{48}F_{3}^{KS}\cdot F_{3}^{KS}\right)h_{ij}(\tau,\bar{x})\right.\label{eq:einsteinqq}\\
\nonumber \end{eqnarray}
where $\bar{x}$ denotes worldvolume coordinates $(x_{1},x_{2},x_{3},x_{4}),\textrm{ $\tau$ is the radial coordinate and }$the
covariant derivative is with respect to the metric \begin{eqnarray}
ds_{10}^{2} & = & h^{-1/2}(\tau)dx_{n}dx_{n}+h^{1/2}(\tau)ds_{6}^{2},\label{eq:KSmetric}\end{eqnarray}
\begin{eqnarray*}
h(\tau)=(g_{s}M\alpha')^{2}2^{2/3}\epsilon^{-8/3}I(\tau) & \,\,\;\,\,\,\, & I(\tau)\equiv\int_{\tau}^{\infty}dx\ \frac{(x\coth x-1)(\sinh(2x)-2x)^{1/3}}{\sinh^{2}x}\end{eqnarray*}
 We will work in the basis $\{\tau,\, g^{1},\, g^{2},\, g^{3},\, g^{4},\, g^{5}\}$
where the deformed conifold metric is diagonal, 

\bea {ds_6}^2=\frac{1}{2}\epsilon^{4/3}K(\tau)\left[\frac{1}{3K^3(\tau)}(d^2\tau+(g^5)^2)\right. &+&\cosh^2(\frac{\tau}{2})[(g^3)^2 +(g^4)^2]\nonumber \\&+& \left.\sinh^2(\frac{\tau}{2})[(g^1)^2 +(g^2)^2]\rule{0cm}{0.6cm}\right]\eea

Note that the dimensionfull parameter $\epsilon$ and $g_{s}M\alpha',$
appear in $h(\tau)$ and in the deformed conifold metric. Define $G_{ij}(\tau)$
such that $g_{ii}(\tau)=h(\tau)^{-1/2}G_{ii}(\tau)$ for $i=1...4$
and $g_{\mu\mu}(\tau)=h(\tau)^{1/2}\epsilon^{4/3}G_{\mu\mu}(\tau)$
for $\mu=5...10$. Explicitly,\begin{eqnarray*}
G_{ii}(\tau)= & \eta_{ii} & i=1...4\\
G_{66}(\tau)= & G_{55}(\tau)= & \frac{1}{6K(\tau)^{2}}\\
G_{88}(\tau)= & G_{77}(\tau)= & \frac{1}{2}K(\tau)\cosh^{2}(\frac{\tau}{2})\\
G_{1010}(\tau)= & G_{99}(\tau)= & \frac{1}{2}K(\tau)\sinh^{2}(\frac{\tau}{2})\end{eqnarray*}

With this notation we obtain for the left hand side of (\ref{eq:einsteinqq}),

\begin{eqnarray}
-\frac{1}{2}\nabla^{\lambda}\nabla_{\lambda}h_{ij}(\tau,\bar{x})\,+\frac{1}{2}\nabla^{l}\nabla_{i}h_{lj}(\tau,\bar{x})\, & + & \frac{1}{2}\nabla^{l}\nabla_{j}h_{li}(\tau,\bar{x})=\frac{e^{ikx}}{g_{s}\alpha'M\sqrt{I(\tau)}G_{55}(\tau)2^{1/3}}\left[-\frac{d^{2}h_{ij}(\tau)}{d\tau^{2}}\right.\nonumber \\
\nonumber \\ & - & \left(\frac{d\ln I(\tau)}{d\tau}+\frac{d\ln G_{99}(\tau)}{d\tau}+\frac{d\ln G_{77}(\tau)}{d\tau}\right)\frac{dh_{ij}(\tau)}{d\tau}\nonumber \\
 & + & \left.\left(\frac{g_{s}^{2}M^{2}\alpha'^{2}G_{55}(\tau)}{2^{1/3}\epsilon^{4/3}}I(\tau)k^{2}-\frac{1}{4I(\tau)^{2}}\left(\frac{dI(\tau)}{d(\tau)}\right)^{2}\right)\right]h_{ij}(\tau)\nonumber \\
\label{eq:explicitricci}\end{eqnarray}
The five form and three form terms are,\begin{eqnarray}
\left(\frac{1}{5}\mathcal{\star F}_{5}^{KS}\cdot\mathcal{\star F}_{5}^{KS}\right)h_{ij}(\tau)=-\frac{e^{ikx}}{g_{s}\alpha'M\sqrt{I(\tau)}G_{55}(\tau)2^{1/3}}\left(\frac{2^{5/3}l(\tau)^{2}}{I(\tau)^{2}K(\tau)^{4}(\cosh^{2}\tau-1)^{2}}\right)h_{ij}(\tau)\nonumber \\
\label{eq:fiveformterm}\end{eqnarray}
\begin{eqnarray}
-\left(\frac{g_{s}^{2}}{48}H_{3}^{KS}\cdot H_{3}^{KS}+\frac{1}{48}F_{3}^{KS}\cdot F_{3}^{KS}\right)h_{ij}(\tau) & = & \frac{e^{ikx}}{g_{s}\alpha'M\sqrt{I(\tau)}G_{55}(\tau)}\nonumber \\
 &  & \left\{ \frac{1}{8I(\tau)}\left[\frac{-1}{G_{77}^{2}(\tau)}\right.\left(\left(\frac{dk(\tau)}{d\tau}\right)+(1\!-F(\tau))^{2}\right)\right.\nonumber \\
 & - & \frac{2}{G_{77}(\tau)G_{99}(\tau)}\left(\left(\frac{dF(\tau)}{d\tau}\right)^{2}+\frac{1}{4}(k(\tau)-f(\tau))^{2}\right)\nonumber \\
 & - & \left.\left.\frac{1}{G_{99}^{2}(\tau)}\left(F(\tau)^{2}+\left(\frac{df(\tau)}{d\tau}\right)^{2}\right)\right]\right\} h_{ij}(\tau)\nonumber \\
\label{eq:threeformterm}\end{eqnarray}
Putting together equations (\ref{eq:explicitricci}),(\ref{eq:fiveformterm})
and (\ref{eq:threeformterm}) we obtain,\[
\frac{d^{2}}{d\tau^{2}}h_{ij}\left(\tau\right)+A\left(\tau\right)\frac{d}{d\tau}h_{ij}\left(\tau\right)+\left(B\left(\tau\right)+\frac{g_{s}^{2}\alpha'^{2}M^{2}}{2^{1/3}\epsilon^{4/3}}I(\tau)G_{55}(\tau)k^{2}\right)h_{ij}\left(\tau\right)=0\]
with\[
A(\tau)=\frac{d\ln(G_{99}(\tau))}{d\tau}+\frac{d\ln(G_{77}(\tau))}{d\tau}+\frac{d\ln I(\tau)}{d\tau}\]
and\begin{eqnarray*}
B(\tau) & = & \frac{-2^{(1/3)}}{8I(\tau)G_{77}(\tau)^{2}}(1-F(\tau))^{2}-(\frac{d}{d\tau}k(\tau))^{2}-\frac{2^{1/3}}{16G_{77}(\tau)G_{99}(\tau)}(k(\tau)-f(\tau))^{2}+\\
 &  & \frac{2^{1/3}}{4}(\frac{d}{d\tau}F(\tau))^{2}-\frac{2^{1/3}}{8I(\tau)G_{99}(\tau)^{2}}((\frac{d}{d\tau}f(\tau))^{2}+F(\tau)^{2})\\
 &  & +\frac{1}{4I(\tau)^{2}}\left(\frac{dI(\tau)}{d\tau}\right)^{2}-\frac{2^{5/3}l(\tau)^{2}}{I(\tau)^{2}K(\tau)^{4}(\cosh^{2}\tau-1)^{2}}.\\
\end{eqnarray*}

The explicit form of $A(\tau)$ and $B(\tau)$ in terms of hyperbolic
functions is obtained by substituting the definitions of section (\ref{sec:KSsugra})
in the expressions above. \vfill\eject\bibliographystyle{utphys}
\bibliography{spin2.bbl}

\end{document}